\documentclass[sigconf]{acmart}

\usepackage{booktabs}

\usepackage{graphicx}
\usepackage{color}
\usepackage{listings}
\usepackage{multirow}
\usepackage{amsmath}
\usepackage{amssymb}
\usepackage{multicol}

\usepackage{labsedc}
\usepackage{amsthm}

\newtheorem{definsec}{Definition}

\definecolor{listingsBackground}{rgb}{0.95,0.95,1.0} 

\lstdefinestyle{xmlCode}{language=XML,frame=tbrl,
  basicstyle=\tiny\scriptsize\ttfamily,keywordstyle=\ttfamily\bf,
  stringstyle=\tiny\itshape,flexiblecolumns=false,
  commentstyle=\itshape,captionpos=b,abovecaptionskip=10pt,belowcaptionskip=0pt,
  backgroundcolor=\color{listingsBackground},frame=none,
  numbers=left,numberstyle=\tiny }

\setcopyright{rightsretained}

\acmDOI{10.475/123_4}

\acmConference[AST2018]{ACM workshop}{May 2018}{Gothenburg, Sweeden}
\acmYear{2018}
\copyrightyear{2018}

\acmArticle{4}
\acmPrice{15.00}

\begin{document}
\title{An automated model-based test oracle for access control systems}

\author{Antonia Bertolino, Said Daoudagh, Francesca Lonetti, Eda Marchetti}
\authornote{Said Daoudagh is also at University of Pisa, Department of Computer Science, Italy.}
    \affiliation{
      \institution{ISTI-CNR}
      \streetaddress{via G.Moruzzi, 1}
      \city{Pisa}
      \state{Italy}
    }
    \email{firstname.secondname@isti.cnr.it}

%
%
%
%
%
%
%
%
%
%
%
%

\begin{abstract}

In the context of XACML-based access control systems,
an intensive testing activity is among the most adopted means to assure that sensible  information or resources are correctly accessed. Unfortunately, it  requires a huge effort for  manual inspection of results: thus automated verdict derivation is a key aspect
for improving the cost-effectiveness of testing.
To this purpose, we introduce XACMET, a novel approach for
automated
model-based oracle definition. XACMET defines a typed graph, called the XAC-Graph, that models  the XACML policy evaluation. The expected verdict of a specific request execution can thus be automatically derived by executing the corresponding path in such graph.
Our validation of the XACMET prototype implementation confirms the effectiveness of the proposed approach.

\end{abstract}

%
%



\copyrightyear{2018}
\acmYear{2018}
\setcopyright{acmlicensed}
\acmConference[AST'18]{AST'18:IEEE/ACM 13th International
Workshop on Automation of Software Test }{May 28--29,
2018}{Gothenburg, Sweden}
\acmBooktitle{AST'18: AST'18:IEEE/ACM 13th International Workshop
on Automation of Software Test , May 28--29, 2018, Gothenburg,
Sweden}
\acmPrice{15.00}
\acmDOI{10.1145/3194733.3194743}
\acmISBN{978-1-4503-5743-2/18/05}

\maketitle

\section{Introduction}
\label{intro}

Security is a primary concern in modern pervasive and interconnected distributed systems.
An important security aspect is constituted by  \textit{access control policies}, which specify which \textit{subjects} can access which \textit{resources} under which \textit{conditions}. They are usually written using  the eXtensible Access Control Markup Language (XACML)~\cite{xacml}, an XML-based standard language proposed by OASIS, and rely on a specific  architecture: incoming access requests are transmitted  to the Policy Decision Point (PDP) that grants or denies the access based on the defined XACML policies.
The criticality of the PDP component, as explained in~\cite{hwang10:policy}, imposes an accurate testing activity that   mainly consists  of  probing the PDP with a set of XACML
requests and checking its responses against the expected decisions.

In literature, there are different proposals for automating PDP testing, including: mutation~\cite{martin07:fault}, coverage~\cite{martin06:defining}, random, combinatorial~\cite{martin06:automated,iet} and
model-based~\cite{xu2015automated} techniques.
However, they share an important drawback: the \emph{lack of oracle}, i.e., for the generated requests the expected PDP decision is not provided. This is an important limitation, especially when test suites are large and manual inspection of results is unfeasible. Recently, Li et al. \cite{multipleImplementation} proposed to implement a PDP automated oracle through voting, i.e. to locally or remotely access more than one PDP engine and collect their responses for the same request.  The most frequent decision value is considered the correct one. Although effective, this solution has a high computation and implementation cost and could not be applied in low energy consuming environments.
Other proposals, for instance \cite{daoudagh2015toolchain}, strictly bind the  oracle definition to the proposed test generation approach and do not provide generic solutions able to evaluate any kind of requests.

In this paper, we introduce XACMET (XACML Modeling \&  Testing), a novel model-based approach
to support the automation of XACML-based testing.
Intuitively, XACMET  builds from the XACML specification a  typed graph, called the XAC-Graph, representing  the XACML policy evaluation.
Such graph can be exploited for several purposes, for example
it allows for measuring the coverage of test requests in terms of the paths executed on the XAC-Graph.
It can also help in deriving an adequate set of test requests such that all paths are executed at least once. However, for lack of space, we focus here
on what is, to the best of our knowledge, the most novel feature supported by XACMET, i.e., \textit{the first
completely automated  model-based oracle for XACML-based PDP testing}. XACMET represents an alternative approach for oracle derivation with respect to the well-known voting mechanism presented in \cite{multipleImplementation}.
We refer to \cite{saidThesis} for the other features of XACMET.

In summary, the contributions of this paper include: i) the definition of the XAC-Graph for modeling XACML policies; ii) the automatic derivation of an XACML oracle based on the evaluation paths of the  XAC-Graph;
and iii) a first empirical evaluation of the XACMET oracle against two different oracle specifications.

The rest of this paper is structured as follows. Section \ref{xacml} briefly introduces XACML.
Section \ref{motivation}  overviews the basic idea for the oracle derivation while
the XAC-graph and the oracle derivation are formally presented in Sections~\ref{def} and~\ref{approach}.  Section~\ref{experiment} reports the validation of the proposal. Finally, Section~\ref{related} puts our work in context of related work and Section~\ref{conclusion} draws conclusions.

\section{XACML}
\label{xacml}

XACML \cite{xacml}\footnote{The current implementation of the presented approach is compliant with XACML 2.0 but it can be easily extended to address the functionalities of XACML 3.0.}
is a platform-independent XML-based standard language for the specification
of access control policies.
Briefly, an XACML policy has a tree structure whose main elements are: PolicySet, Policy, Rule, Target and Condition. The PolicySet includes one or more
policies. A Policy contains a Target and one or more rules. The Target specifies a
set of constraints on attributes of a given request. The Rule specifies a Target and a Condition containing one
or more boolean functions.
If a request satisfies the target of the policy,
then the set of rules of the policy is checked, else the policy is skipped.
If the Condition evaluates to true, then the Rule's
Effect (a value of Permit or Deny) is returned, otherwise a NotApplicable decision
is formulated (Indeterminate is returned in case of errors).
More policies in a policy set and more rules in a policy may be applicable to a given request.
The PolicyCombiningAlgorithm and the RuleCombiningAlgorithm define how to combine the results from multiple policies and rules respectively in order to derive a single access result.

For example, the \emph{first-applicable} rule combining algorithm returns the effect of the first applicable rule or \emph{NotApplicable} if no rule is applicable to the request.
The \emph{deny-overrides} algorithm specifies that \emph{Deny} takes the precedence
regardless of the result of evaluating any of the other rules in the combination, then
it returns \emph{Deny} if there is a rule that is evaluated to \emph{Deny}, otherwise it returns \emph{Permit}
if there is at least a rule that is evaluated to \emph{Permit} and all other rules are evaluated to \emph{NotApplicable}. Similarly, the \emph{permit-overrides} algorithm returns \emph{Permit} if there is a rule that is evaluated to \emph{Permit}.

\begin{lstlisting}[style=xmlCode, breaklines,
caption={An XACML
policy},label={lst:policy-example}]
<Policy PolicyId="Pol_Ex" RuleCombiningAlgId="deny-overrides">
 <Target/>
 <Rule RuleId="ruleA" Effect="Deny">
  <Target>
   <Resources><Resource><ResourceMatch MatchId="string-equal">
     <AttributeValue DataType="string">book</AttributeValue>
     <ResourceAttributeDesignator AttributeId="resource-id" DataType="string"/>
    </ResourceMatch></Resource>
    <Resource><ResourceMatch MatchId="string-equal">
     <AttributeValue DataType="string">document</AttributeValue>
     <ResourceAttributeDesignator AttributeId="resource-id" DataType="string"/>
    </ResourceMatch></Resource>
    <Resource><ResourceMatch MatchId="string-equal">
     <AttributeValue DataType="string">documententry</AttributeValue>
     <ResourceAttributeDesignator AttributeId="resource-id" DataType="string"/>
    </ResourceMatch></Resource></Resources>
   <Actions><Action><ActionMatch MatchId="string-equal">
     <AttributeValue DataType="string">write</AttributeValue>
     <ActionAttributeDesignator AttributeId="action-id" DataType="string"/>
    </ActionMatch></Action></Actions>
  </Target>
  <Condition><Apply FunctionId="string-is-in">
    <Apply FunctionId="string-one-and-only">
     <ResourceAttributeDesignator AttributeId="resource-id" DataType="string"/>
    </Apply>
    <SubjectAttributeDesignator AttributeId="subject-id1" DataType="string"/>
   </Apply>
  </Condition>
 </Rule>
 <Rule RuleId="ruleB" Effect="Permit">
  <Target>
   <Subjects><Subject><SubjectMatch MatchId="string-equal">
     <AttributeValue DataType="string">Julius</AttributeValue>
     <SubjectAttributeDesignator AttributeId="subject-id" DataType="string"/>
    </SubjectMatch></Subject></Subjects>
   <Resources><Resource><ResourceMatch MatchId="string-equal">
     <AttributeValue DataType="string">journals</AttributeValue>
     <ResourceAttributeDesignator AttributeId="resource-id" DataType="string"/>
    </ResourceMatch></Resource></Resources>
   <Actions><Action><ActionMatch MatchId="string-equal">
     <AttributeValue DataType="string">read</AttributeValue>
     <ActionAttributeDesignator AttributeId="action-id" DataType="string"/>
    </ActionMatch></Action></Actions>
  </Target>
 </Rule>
</Policy>
\end{lstlisting}

%
%
%
%
%

We show in Listing \ref{lst:policy-example} an example of a simplified XACML
policy ruling library access. Its target (line 2) says that this policy
applies to any subject, resource and action. This policy has a first rule, \emph{ruleA}
(lines 3-29), with a target (lines 4-21) specifying that this rule applies only
to the access requests of a ``write'' action of ``book'', ``document'' and ``documententry''
resources. The rule condition will be evaluated true when the request resource value is contained into the set of request subject values.
The effect of the second rule \emph{ruleB}
(lines 30-45) is \emph{Permit} when the subject is ``Julius'', the action is ``read'', and the resource is
``journals''. The rule combining algorithm of the
policy (line 1) is \emph{deny -overrides}.

\section{Test oracle}
\label{motivation}

By referring to the example policy of Listing \ref{lst:policy-example}, in this section we explain the underling idea of the approach used for the oracle definition. We refer to  Section~\ref{def} for formal details.

Given a generic request, the  result of the evaluation of an XACML policy with that request strictly depends on: the request values, the policy constraints as well as the combining algorithm that prioritizes the evaluation of the policy rules.
Specifically, we define an \emph{evaluation path} as a sequence of policy elements that are  exercised by the request during the evaluation of an XACML policy and the verdict associated to that request.
Thus, the general idea of the XACMET approach is to derive all possible evaluation paths\anto{ma non e' stato detto  cosa sia un evaluation path} from the policy specification and order them according to the rule combining algorithm.
For instance, let us consider the policy of Listing \ref{lst:policy-example}, having as elements, the rules \emph{ruleA} and \emph{ruleB} and \emph{deny-overrides} as the combining algorithm (line 1), the possible evaluation paths are: (1) \emph{ruleA} evaluated to true and  \emph{ruleB} evaluated to false: the associated verdict is \emph{Deny}, i.e., the effect of the first rule; (2) \emph{ruleA} evaluated to false and \emph{ruleB} evaluated to true: the associated verdict is \emph{Permit}, i.e., the effect of the second rule; (3) \emph{ruleA} and \emph{ruleB} both evaluated to false: the associated verdict is \emph{NotApplicable}; (4)  \emph{ruleA} and \emph{ruleB} both evaluated to true: the associated verdict is \emph{Deny}, because it takes the precedence regardless of the result of the second rule.
Note that, for aim of simplicity, we do not explicitly consider the \emph{Indeterminate} value that is returned in case of errors of the policy evaluation, and we assume that in this case the associated verdict is \emph{NotApplicable}.
This set of paths is ordered according to the semantics of the rule combining algorithm, and then according to the verdict associated to each path.
For instance, in case of \emph{deny-overrides} combining algorithm, first the paths having \emph{Deny} are evaluated, then those having \emph{Permit} and finally those having \emph{NotApplicable}.
For paths having the same verdict, the evaluation order of the paths is based on their length, namely the shortest path takes the precedence.
For the policy of Listing \ref{lst:policy-example}, the order of the evaluated paths is (1), (4), (2) and (3).

The ordered set of paths is then used for the requests evaluation and the verdicts association. For each request, the first path for which all the path constraints are satisfied by the request values is identified, the final verdict associated to the request is derived, and the path  is considered covered by the request.
For instance, considering the ordered set of paths of policy of Listing \ref{lst:policy-example}, a request asking to \texttt{write} a \texttt{documententry} does not match paths (2), (3) and (4), but covers (satisfies) path (1) since it satisfies only \emph{ruleA}, so the associated verdict is \emph{Deny}.

In addition, XACMET also provides the possibility of automatically generating a set of test cases that guarantee the full coverage of the evaluation paths.
Indeed, each of the evaluation path represents the set of constraints that should  be satisfied by some specific request values so to reach the final verdict. Thus, the set of values  satisfying (not satisfying) the identified set of path constraints can be identified using  a constraint satisfaction approach \cite{turkmen2015analysis}.
As a side effect, when  the various constraints are combined, possible inconsistencies between the selected values can be detected, which hints at the potential presence of unfeasible paths in the policy specification.
Moreover, XACMET application gives the possibility of knowing which and how many evaluation paths are covered by a test set.
This information can be useful to improve the policy itself and avoid possible security flaws. For space limitation, the test cases generation functionality of XACMET is not further described in this paper. We refer to \cite{saidThesis} for more details.

\section{XAC-graph}
\label{def}

\begin{figure*}[]
\centering
\includegraphics[scale=0.23]{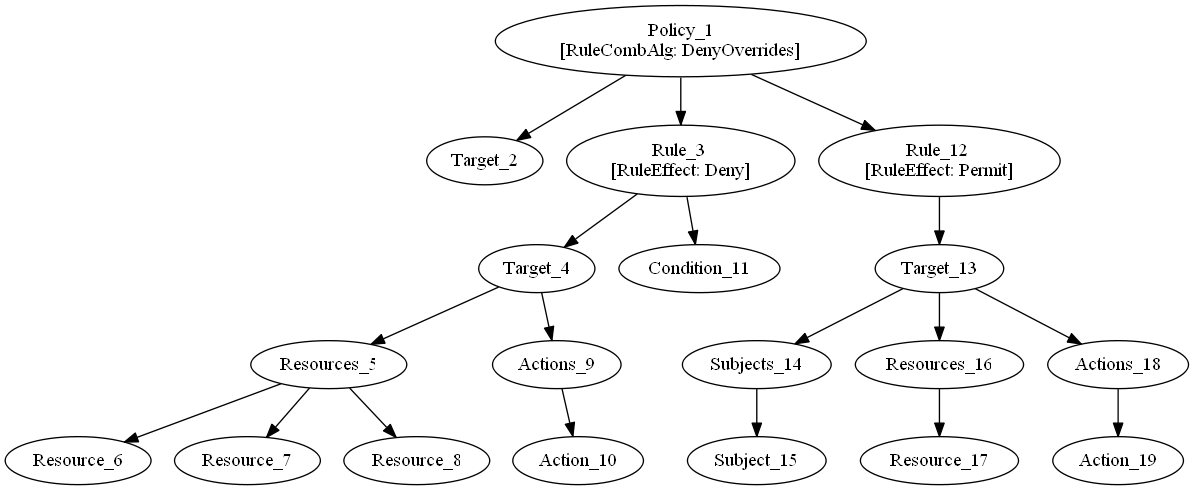}
\caption{XAC-Tree. Label T\_P means node of type T and parameter P. The attributes are within square brackets.} \label{fig:tree}
\end{figure*}

In this section, we provide some formal definitions related to the
XAC-graph model.

With reference to the XACML policy,
as an XML document it can be represented as a tree, called the XAC-Tree. In particular,  the following concepts can be used:

\begin{itemize}
  \item \emph{Contained:} Element i is contained within element j if i is between the start-tag and the end-tag of  j.
  \item\emph{Parent:} Element i is the parent of element j when j is contained within i and i  is exactly one level above j.
  \item \emph{Sibling}:  The siblings in an XML document are the elements that are on a same level of the tree and have the same parent. In particular, given parent i of elements j and k, j is left (right) sibling  of k if j is contained just before (after) k within element i.
      \end{itemize}

The XAC-Tree derivation exploits the  parent relationship of the XACML policy and uses the following sets of types and values:
\begin{itemize}
  \item $T_{V}$ = \{Policy, Target Rule, Subjects, Subject, Resources, Resource, Actions, Action, Environments, Environment\};
  \item $T_{V_{a}}$  = \{RuleAlgorithm, Effect, NotApplicable\};
  \item $T_{V_{v}}$ =\{ReturnPermit, ReturnDeny, ReturnNotApplicable\};
  \item \texttt{RCA}= \{FirstApplicable, DenyOverrides, PermitOverrides\};
  \item \texttt{RE} = \{Permit, Deny\}.
\end{itemize}

Formally, the parent relationship, called XAC-TreeParent, is defined as:

\begin{definsec}[XAC-TreeParent]
\label{lab:XtreeParent}

Given a tree T=(V, E, Root) in which every vertex v in V has an associated type $t_{v} \in T_{V}$: Element i $\in$ \emph{V} is  \emph{XAC-TreeParent} of j $\in$ V if i is parent of j in the XACML policy document.
\end{definsec}

From this, the definition of the XAC-Tree as in the following:

\begin{definsec}[XAC-Tree]
\label{lab:Xtree}
Given an XACML policy,
\anto{questo file e' lo stesso di document sopra? userei sempre la stessa terminologia}\fra{fatto}
the XAC-Tree is  a labeled and typed tree (V, E, Root) where
\begin{itemize}
\item V is a set of vertices such that every  v $\in$ V has type $t_{v} \in T_{V}$;
\item Each vertex v in V has parameter i  with i = 1,.., n;
\item E is a set of edges (i,j) such that i,j $\subseteq$ V and i is \emph{XAC-TreeParent} of j;
\item Root is a vertex v in V with $t_{Root}$ = Policy;
\item Root has an attribute called \emph{RuleCombAlg }$\in$ RCA;
\item Each vertex v $\in$ V with  $t_{v}$ = Rule  has  an attribute called \emph{EffectRule} $\in$ RE.

\end{itemize}
\end{definsec}

Figure \ref{fig:tree} shows the  XAC-Tree associated to  the policy of Listing \ref{lst:policy-example}. In particular, \emph{ruleA}  becomes the node \texttt{Rule\_3} into the XAC-Tree. In this case, 3 is the suffix of the node and the  EffectRule attribute of the node \texttt{Rule\_3}  is set to Deny as specified in the Effect of \emph{ruleA}.
Moreover, there is a XAC-TreeParent relation between \texttt{Policy\_1} and \texttt{Rule\_3} nodes because \emph{ruleA} is contained within \texttt{policyExample} in Listing \ref{lst:policy-example}.

The representation of the XACML policy is then used  to derive a model of the XACML evaluation. For this, a parent relationships, called XAC-GraphParent, is defined as in the following:

\begin{definsec}[XAC-GraphParent]
\label{lab:graphparent}

Given a graph G= ($V_{g}$,$E_{g}$,Entry) and XT = (V,E,Root) a XAC-Tree, where Entry = Root and
$V_{g}= V \bigcup V_{a}^{\star}\bigcup V_{v}^{\star}$ with (i) $V_{a}$ is a set of vertices such that every \emph{v} $\in V_{a}$ has type $t_{v} \in T_{V_{a}}$; (ii) $V_{v}$ is a set of vertices such that every \emph{v} $\in V_{v}$ has type $t_{v} \in T_{V_{v}}$.

Element i $\in$ $V_{g}$ is in a \emph{XAC-GraphParent} relation with  j $\in$ $V_{g}$ if:

\begin{enumerate}
\item  i is left sibling of j in XT, $t_{j}$ $\neq$ Rule and i has not children in XT;
\item  i is leaf in XT and $\exists$ k $\in$ $V_{g}$ such that k is XAC-TreeParent of i and k is left sibling of j in XT and $t_{j}$ $\neq$ Rule;
\item  j has no left sibling in XT and i is XAC-TreeParent of j in XT;
\item $t_{j}$ $\in$ \{Effect, NotApplicable\}, i is leaf in XT,  $t_{i}$ $\neq$ Target,  $\exists$ k $\in$ V such that $t_{k}$=Rule and i is the rightmost  leaf of the subtree rooted in k in XT;
\item $t_{i}$ $\in$ \{Effect, NotApplicable\} and  $t_{j}$ = RuleAlgorithm;
\item $t_{i}$= RuleAlgorithm, $t_{j}$ = ReturnPermit if $\exists$ k$\in $ V, $t_{k}$=Rule such that the value of the attribute EffectRule of k is Permit;
\item $t_{i}$= RuleAlgorithm,  $t_{j}$ = ReturnDeny if $\exists$ k$\in $ V, $t_{k}$=Rule such that the value of the attribute EffectRule of k is Deny;
\item $t_{i}$= RuleAlgorithm and  $t_{j}$ = Rule, and if k $\in$ V is the left sibling of j in XT than $t_{k}$$\neq$ Target.
\end{enumerate}
\end{definsec}

Finally, a labelled and typed graph, called the XAC-Graph,  is defined as in the following:

\begin{definsec}[XAC-Graph]
\label{lab:XGRAPH}
Let XT = (V,E,Root) be a XAC-Tree, a policy graph (XAC-Graph) is a graph ($V_{g}$,$E_{g}$,Entry) where \begin{itemize}
\item Entry = Root;
\item $V_{g}= V \bigcup V_{a}^{\star}\bigcup V_{v}^{\star}$
\item $V_{a}$ is a set of vertices such that every v $\in V_{a}$ has type $t_{v} \in T_{V_{a}}$;
\item $V_{v}$ is a set of vertices such that every v $\in V_{v}$ has type $t_{v} \in T_{V_{v}}$;
\item Each vertex v in $V_{a}$ with $t_{v}$ = Effect has an attribute called \emph{EffectValue} $\in$ RE;
\item Each vertex v in $V_{a}$ has parameter ${i}$  with ${i}$ = 1,.., n;
\item The vertex v in $V_{a}$ with $t_{v}$ = RuleAlgorithm is unique and has an attribute called \emph{Algorithm} of value equal to the value of the attribute RuleCombAlg of the Root in XT;
\item E is a set of edges (i,j) such that i,j $\sqsubseteq$ $V_{g}$ and i is XAC-GraphParent of j;
\item Each vertex j, j, k $\in$ $V_{g}$ with $t_{j}$ = Effect, i XAC-GraphParent for the point 4 of the definition \ref{lab:graphparent}  and $t_{k}$=Rule the value of the attribute EffectValue of j is equal to value of the attribute EffectRule of k.
\end{itemize}
\end{definsec}

\begin{figure}[]
\centering
\includegraphics[scale=0.22]{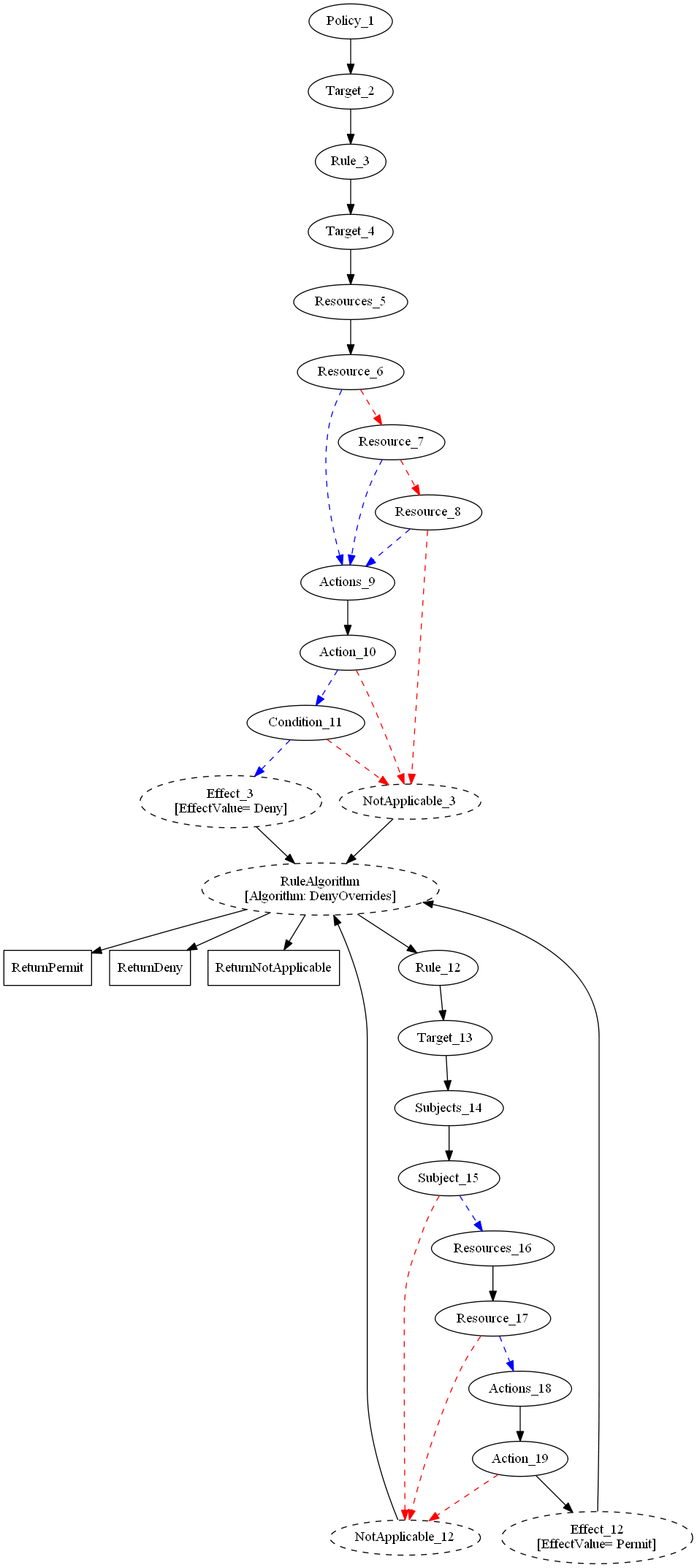}
\caption{XAC-Graph. Label T\_P means  node of type T and parameter P. The attributes are within square brackets.}
\label{fig:GRAPH}
\end{figure}

Figure \ref{fig:GRAPH} shows the  XAC-Graph of the  XAC-Tree of Figure \ref{fig:tree}.
In particular, considering the conditions of Definition \ref{lab:graphparent}, in XAC-Graph there exists a  XAC-GraphParent relation between:
\begin{itemize}
  \item  \texttt{Target\_2} and \texttt{Rule\_3} due to Condition 1;
  \item  \texttt{Resource\_7} and \texttt{Actions\_9} due to Condition 2;
  \item  \texttt{Policy\_1} and \texttt{Target\_2} due to Condition 3;
  \item  \texttt{Condition\_11} and \texttt{NotApplicable\_3} due to Condition 4;
  \item  \texttt{Condition\_11} and \texttt{Effect\_3} due to Condition 4;
  \item  \texttt{Effect\_3} and \texttt{RuleAlgorithm} due to Condition 5. Note that \texttt{Policy1} and \texttt{RuleAlgorithm} have the same algorithm value;
  \item  \texttt{RuleAlgorithm} and \texttt{ReturnPermit} due to Condition 6;
  \item  \texttt{RuleAlgorithm} and \texttt{ReturnDeny} due to Condition 7;
  \item  \texttt{RuleAlgorithm} and \texttt{Rule\_12} due to Condition 8.
\end{itemize}

The XAC-graph could basically be derived applying a depth-first search approach to the XAC-tree.
For the sake of simplicity, we show the XACMET approach applied to a policy rooted in the Policy element. Definitions \ref{lab:XtreeParent}, \ref{lab:Xtree}, \ref{lab:graphparent}, \ref{lab:XGRAPH},  can be easily extended to also consider the \emph{PolicySet} and the XACMET approach can be used also for deriving the XAC-Graph considering  a policy rooted in the \emph{PolicySet} node.

\section{Oracle derivation}
\label{approach}
In this section, we detail the approach used for the oracle derivation that is based on the XAC-Graph
paths identification. Specifically, the process adopted is divided into two main steps: \emph{coloring, unfolding}.

The coloring process exploits the concepts of \emph{Forward Node} defined as: given the XAC-Graph
 G = ($V_{g}$ , $E_{g}$, Entry), for each node i $\in$ $V_{g}$ it is possible to identity
 the Forward Node $FN(i)$ $\in$ $V_{g}$ as the set of nodes j such that i is a XAC-GraphParent of j.
Thus, given a XAC-Graph for each node b and c  $\in$ $V_{g}$,  with  $t_{b}$ $\in$ \{Subject, Resource, Action, Environment\}, the cardinality of $FN(b)$ = 2, and $t_{c}\in$
\{Subject, Resource, Action, Environment, NotApplicable\},  the coloring process marks each edge (b,c)$\in$ $FN(b)$: with red dashed line if $t_{b}=t_{c}$  or $t_{c}$ = NotApplicable; with blue dotted line otherwise.
In practice the red dashed edges represent a successful evaluation of the node b.

During the unfolding process the paths  are obtained by visiting the XAC-Graph from the Entry node to each node in $V_{v}$. The cycles are due to the presence of the node typed \texttt{Rule\_Algorithm}. In XACMET, the order of the paths strictly depends on the order in which the rules are evaluated, which in turn is guided by  the  FirstApplicable, DenyOverrides, PermitOverrides  algorithms.

Thus, let P a path of k nodes on the XAC-Graph and h the last included one, with  $t_{h}$ =Rule\_Algorithm. If the value of Algorithm attribute of h is equal to:

\textbf{FirstApplicable and the node k-1 has type:}

\begin{itemize}
  \item  Effect, and EffectValue = Deny (Permit), then the next node has type ReturnDeny (ReturnPermit);
  \item NotApplicable, then the next node v has type Rule iff v is not already included in P (ReturnNotApplicable otherwise).
\end{itemize}

\textbf{DenyOverrides and the node k-1 has type}

 \begin{itemize}
   \item  Effect and EffectValue = Deny, then the next node has type ReturnDeny;
   \item  Effect and EffectValue = Permit, then the next node v has type Rule iff v is not already included in P (ReturnPermit otherwise);
  \item NotApplicable, then the next node v has type Rule iff v is not already included in P;
 \item  NotApplicable and each node v with Rule $\in P$ the next node has type ReturnPermit, if $\exists$ a node p $\in$ $P$ with  $t_{p}$=Effect (ReturnNotApplicable otherwise).
\end{itemize}

\textbf{Permit\_Overrided and the node k-1 has type}

\begin{figure}[]
\centering
\includegraphics[scale=0.28]{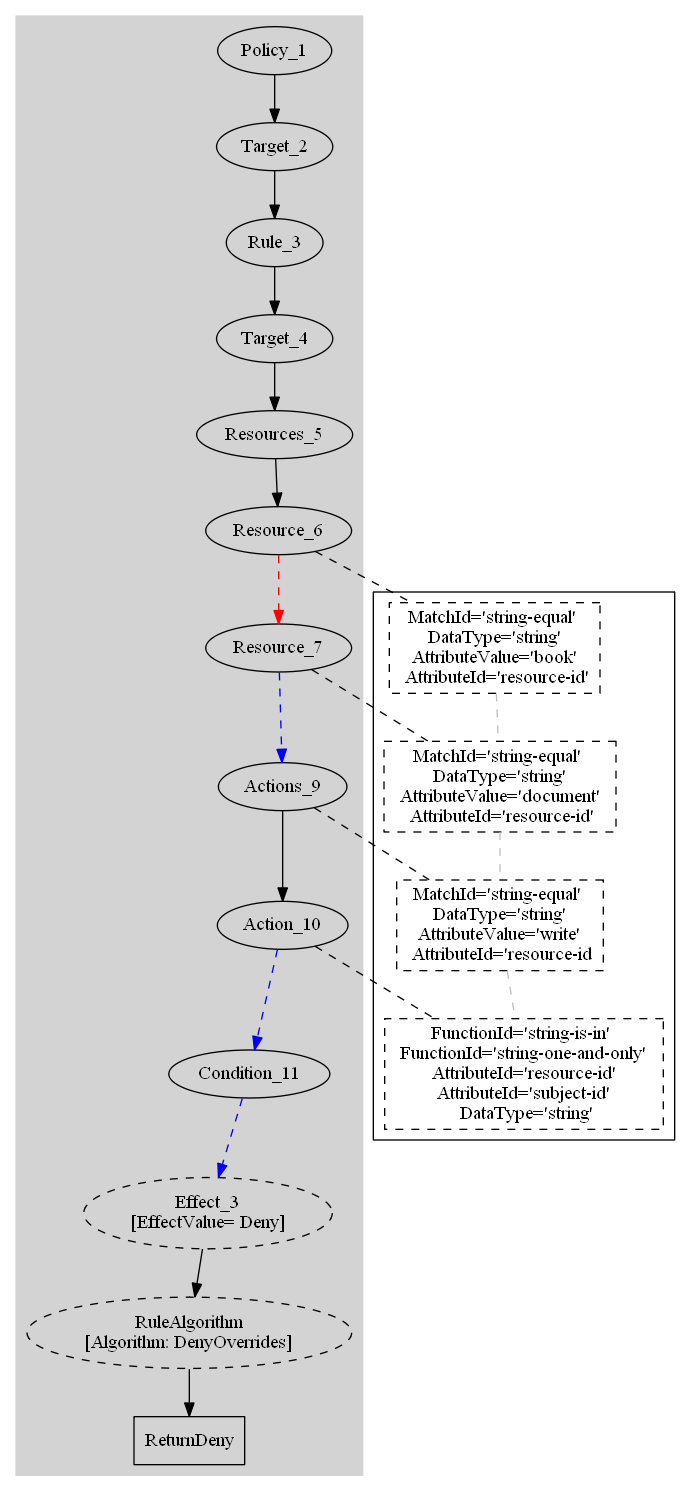}
\caption{A path of XAC-Graph. The boxes connected to the nodes contain the functions and values.} \label{fig:XPATH}
\end{figure}

 \begin{itemize}

   \item  Effect and EffectValue = Permit, then the next node has type ReturnPermit;
   \item  Effect and EffectValue = Deny, then the next node v has type Rule iff v is not already included in P (ReturnDeny otherwise);
   \item NotApplicable, then the next node v has type Rule iff v is not already included in P;
 \item NotApplicable and each node v with Rule $\in P$  the next node has type ReturnDeny, if $\exists$ a node p $\in$ $P$ with  $t_{p}$=Effect (ReturnNotApplicable otherwise).
\end{itemize}
In Figure \ref{fig:XPATH}, we show a path of XAC-Graph.

\section{Empirical evaluation}
\label{experiment}
We conducted an empirical evaluation of the XACMET oracle.
The evaluation includes two studies: a first study assessed the compliance of the XACMET oracle with the XACML specification given by the conformance test suite; the focus of the second study was to assess the compliance of the XACMET oracle with respect to a voting PDP mechanism, considering both the policies of conformance test suite and some real policies.

\subsection{First Study}

In this first study, we considered the tests of the XACML 2.0 Conformance Tests V0.4 \cite{conformancexacml}. Each test consists of three elements: an XACML policy, an XACML request, and an XACML response representing the expected access decision associated to that request. We focused on the subset of tests implementing the mandatory functionalities and specifically on the following groups of tests: i) IIA: tests exercising attribute referencing; ii) IIB: tests exercising target matching; iii) IIC: test exercising function evaluation; and finally iv) IID: tests exercising combining algorithms. For each group, we selected a subset of tests specifying only the functionalities implemented in XACMET. In particular, we excluded by IIA group the tests referring to \emph{Indeterminate} values, in IIC group we considered only the most used arithmetic and equality functions (for instance \emph{type-equal}) while for IID group we considered the combining algorithms presented in Section \ref{approach}.
In Table \ref{tab:PoliciesSubjects} (column 1),  we show for each group the percentage of considered tests. Moreover, in rows from 4 to 7, the structure of the considered XACML policies belonging to the four groups is described in terms of cardinality of policies, rules, conditions, subjects, resources, actions and functions (total number of functions and distinct functions) respectively.
Table \ref{tab:PoliciesSubjects} (column 9) shows the number of XACML requests for the four groups of tests.
To validate the XACMET oracle, we applied the XACMET approach to the policies of the conformance tests.
Specifically, for each test case we derived starting from the XACML policy, the associated XAC-Graph and an ordered set of paths as described in Section \ref{def} and \ref{approach}. Then, we evaluated the XACML request belonging to the test case, against the obtained set of paths, we identified the first covered path and derived the verdict associated to that path. Finally, we compared this verdict with the decision value specified in the response belonging to the test case.
For all the tests of the conformance, we obtained that the XACMET verdict coincides with the expected access decision.
This improves the confidence in the effectiveness of the XACMET approach for the XACML functionalities specified in the conformance tests.

\begin{table*}[ht]
\caption{Experimental values} \label{tab:PoliciesSubjects}
\begin{center}
\resizebox{0.67\textwidth}{!}{
\begin{tabular}{llllllllll}
\toprule
XACML Policy & \multicolumn{7}{c}{Functionality} & \multicolumn{2}{c}{\# XACML Request}\\
    \cmidrule{2-8}
  & \# Policy & \# Rule & \# Cond & \# Sub & \# Res & \# Act & \# Funct (distinct) & I Study & II Study \\

\midrule
\multicolumn{10}{c}{\emph{Conformance Test Suite XACML Policies}}\\
\midrule
IIA (90\%) & 18 & 18  & 12  & 18  & 8  & 16  & 112 (12) &18 & 68  \\
IIB (100\%) & 53 & 53  & 6  & 51  & 50  & 98  & 410 (7) &53 & 254  \\
IIC (10\%) & 22 & 22  & 22  & 18  & 3  & 1  & 102 (19) & 22 & 31\\
IID (17\%) & 5 & 13  & 7  & 13  & -  & -  & 60 (5) & 5 & 19\\
\midrule
\multicolumn{10}{c}{\emph{Real-world XACML Policies}}\\
\midrule
2\_73020419964\_2  & 1& 6  & 5  & 3  & 3  & 0  & 4 & - & 8\\
create-document-policy & 1& 3  & 2  & 1  & 2  & 1  & 3 & - & 5\\
demo-5   & 1& 3  & 2  & 2  & 3  & 2  & 4 & - & 13\\
demo-11   & 1& 3  & 2  & 2  & 3  & 1  & 5 & - & 8\\
demo-26   & 1& 2  & 1  & 1  & 3  & 1  & 4 & - & 16\\
read-document-policy  & 1& 4  & 3  & 2  & 4  & 1  & 3 & - & 6\\
read-informationunit-policy  & 1& 2  & 1  & 0  & 2  & 1  & 2 & - & 4\\
read-patient-policy  & 1& 4  & 3  & 2  & 4  & 1  & 3 & - & 6\\
Xacml-Nottingham-Policy-1  & 1& 3  & 0  & 24  & 3  & 3  & 2 & - & 18\\

\bottomrule

\end{tabular}
}
\end{center}
\end{table*}

\subsection{Second Study}
The goal of this second study was to evaluate the XACMET oracle with respect to other competing automated oracles.
We referred to the already mentioned black box approach of multiple implementations testing presented in \cite{multipleImplementation}. Hence we derived an automated majority oracle
by running the request on three implementations of an XACML based PDP and then using a majority voting to derive the expected XACML response associated to that request.
Specifically, we used
a pool of three XACML PDPs, namely Sun PDP \footnote{Sun PDP is available at: \url{http://sunxacml.sourceforge.net}.}, Herasaf PDP \footnote{Herasaf PDP is available at: \url{https://bitbucket.org/herasaf/herasaf-xacml-core}.} and Balana PDP \footnote{Balana PDP is available at: \url{https://github.com/wso2/balana}.}.

In this second study,
we used XACMET to derive starting from a XACML policy a set of XACML requests and the associated verdict as described in Section \ref{motivation}.
The considered policies were: i) those of the conformance tests described before; ii) the Fedora (Flexible Extensible Digital Object Repository Architecture) XACML policies (demo-5, demo-11 and demo-26) \footnote{Fedora Commons Repository Software. http://fedora-commons.org.} \anto{come sono state scelte? c'e'un criterio che possiamo spiegare - lo chiedevano anche la prima volta che abbiamo sottomesso}
\fra{fatto} and  other six policies released in the context of the TAS3 European
project \footnote{Trusted Architecture for Securely Shared Services. http://www.tas3.eu.}.\anto{mettere reference}
\fra{fatto}
For each policy, Table \ref{tab:PoliciesSubjects} (column 10) shows the cardinality of the derived XACML requests.
Each policy and the derived set of requests were executed on the PDPs pool, we observed whether the different PDPs  produced the same responses and took as oracle value the majority output.
For each request we compared the XACMET oracle with the majority output of the PDPs pool.
We observed that for all requests the XACMET oracle coincided with the automated majority oracle.
This enhances the confidence of the effectiveness of the XACMET oracle considering also the functionalities of real policies.
Moreover, in this second study we also applied the test cases generation functionalities of XACMET. It is out of scope of this paper to validate them and we refer to \cite{saidThesis} for an experimental evaluation of XACMET test suites.
However, as side effect of this experiment we obtained an enhancement of the requests
of the conformance tests. As showed in Table \ref{tab:PoliciesSubjects} (column 10), this enhancement is evident mostly for IIB policies group (201 additional requests). This is due to the more complex structure of the policies of this group as evidenced also by the higher number of elements (mainly actions and functions) of these policies.

\section{Related work}
\label{related}

The work presented in this paper spans over the following research
directions:

\paragraph{Analysis and modeling of policy specification}

Available proposals include different verification techniques \cite{xu2014specification}, such as model-checking~\cite{Zhang} or SAT solvers~\cite{turkmen2015analysis}.
Well-known analysis and verification tools for access control policies are: i) Margrave~\cite{fisler2005verification}, which represents policies as Multi-Terminal Binary Decision Diagrams (MTBDDs) and can answer queries about policy properties; and ii) ACPT (Access Control Policy Testing) tool \cite{hwang2010acpt} that transform policies into finite state machines and represent static, dynamic and historic
constraints into Computational Tree Logic.
The capabilities and performances of such tools are analytically evaluated in \cite{li2015evaluating}.

Differently from the above approaches, XACMET models the expected behaviour of the evaluation
of a given XACML policy as a labeled graph and guarantees the full path coverage of such graph.
Moreover, our proposed XAC-Graph model is richer since it also represents the rule combining algorithm, the functions, and the associated conditions.
The authors of \cite{pina2012graph} provide an optimized approach for XACML policies modeling based
on tree structures aimed at fast searching and evaluation of applicable rules.
Differently from our proposal, the main focus of this work is on performance optimization  more than on oracle derivation.

\paragraph{Test cases and oracle derivation}

Considering the automated test cases generation, solutions have
been proposed for testing either the XACML policy or the PDP
implementation \cite{bertolino2014testing,iet}. Among them, the most referred ones such as X-CREATE and the
Targen tools \cite{martin06:automated,iet} use combinatorial approaches for
test cases generation. 
However, combinatorial approaches are shallow with respect to policy semantics.

Model-based testing has already been widely investigated for policy testing, e.g.~\cite{pretschner2008model,xu2015automated}.
Such approaches provide methodologies or tools for automatically
generating access control test models from functional
models and access control rules.
The key original aspect of our approach is in the XAC-Graph model which we derive,
which is richer in expressiveness than other proposed models, and can provide directly the evaluation paths including a verdict associated to a request.
%
About the automated oracle, notwithstanding the huge interest devoted
to this topic, reducing the human activity in the evaluation of the testing
results is still an issue \cite{barr2015oracle}.
The automated oracle derivation is a key aspect in the context of XACML systems and testers need usually
to manually verify the XACML responses. The few available solutions mainly deal with model based approaches. Specifically, the authors of \cite{daoudagh2015toolchain} provide an integrated toolchain including test
case generation as well as policy and oracle specification
for the PDP testing.
Other proposals such as \cite{telerise2017} address the use of monitoring facilities for the assessment of the run-time execution of XACML policies.
Differently from the above approaches, the main benefits of XACMET deal with the derivation of an XACML verdict for each XACML request.

\section{Conclusions}
\label{conclusion}

We have introduced a novel model-based approach to automatic generation of XACML oracle for testing policy evaluation engines.
The XACMET approach fully automatically derives a verdict for each XACML request by considering the set of evaluation paths derived from the obtained graph. We have illustrated the approach on an example policy and provided experimental results evidencing the effectiveness of our proposal with respect to the oracle provided in the XACML conformance tests and an automated oracle implemented as a voting mechanism.

In the future, we plan to extend our automated oracle in order to consider more functionalities of the XACML conformance policies, such as the different combining algorithms and the \emph{PolicySet} element. The XACMET approach will also be extended to be compliant with the last XACML 3.0 standard version.
Future work will also include further experimentation of XACMET, and its comparison with other model-based approaches.

\section*{Acknowledgments}

This work has been partially supported by the GAUSS national research
project (MIUR, PRIN 2015, Contract 2015KWREMX).

\bibliographystyle{ACM-Reference-Format}
\bibliography{biblio}

\end{document}